\documentclass[11pt]{article}
\usepackage{amssymb}

\title{The Paraconsistent Approach to\\Quantum Superpositions Reloaded:\\Formalizing Contradictiory Powers\\ in the Potential Realm}

\author{{\sc N. C. A. da Costa}$^{1}$\ and {\sc C. de Ronde}\ $^{2}$}
\date{}

\begin{document}

\bibliographystyle{plain}
\maketitle

\begin{center}
\begin{small}
1. Federal University of Santa Catarina - Brazil \\
2. Philosophy Institute ``Dr. A. Korn" \\ 
University of Buenos Aires, CONICET - Argentina \\
Center Leo Apostel and Foundations of  the Exact Sciences\\
Brussels Free University - Belgium \\
\end{small}
\end{center}

\begin{abstract}
\noindent In \cite{daCostadeRonde13} the authors of this paper argued in favor of the possibility to consider a Paraconsistent Approach to Quantum Superpositions (PAQS). We claimed that, even though most interpretations of quantum mechanics (QM) attempt to escape contradictions, there are many hints ---coming from present technical and experimental developments in QM--- that indicate it could be worth while to engage in a research of this kind. Recently, Arenhart and Krause have raised several arguments against the PAQS \cite{ArenhartKrause14a, ArenhartKrause14b, ArenhartKrause14c}. In \cite{deRonde14, deRonde15a} it was argued that their reasoning presupposes a metaphysical stance according to which the physical representation of reality must be exclusively considered in terms of the equation: Actuality = Reality. However, from a different metaphysical standpoint their problems disappear. It was also argued that, if we accept the idea that quantum superpositions exist in a (contradictory) potential realm, it makes perfect sense to develop QM in terms of a paraconsistent approach and claim that quantum superpositions are contradictory, contextual existents. Following these ideas, and taking as a standpoint an interpretation in terms of the physical notions of {\it power} and {\it potentia} put forward in \cite{deRonde13, deRonde15a, deRonde15d}, we present a paraconsistent formalization of quantum superpositions that attempts to capture the main features of QM. 
\end{abstract}

\begin{small}


\end{small}

\bibliography{pom}

\newtheorem{theo}{Theorem}[section]

\newtheorem{definition}[theo]{Definition}

\newtheorem{lem}[theo]{Lemma}

\newtheorem{met}[theo]{Method}

\newtheorem{prop}[theo]{Proposition}

\newtheorem{coro}[theo]{Corollary}

\newtheorem{exam}[theo]{Example}

\newtheorem{rema}[theo]{Remark}{\hspace*{4mm}}

\newtheorem{example}[theo]{Example}

\newcommand{\proof}{\noindent {\em Proof:\/}{\hspace*{4mm}}}

\newcommand{\qed}{\hfill$\Box$}

\newcommand{\ninv}{\mathord{\sim}} %involutive negation

\newtheorem{postulate}[theo]{Postulate}

\section*{Introduction}

In \cite{daCostadeRonde13}, the authors of this paper argued in favor of the possibility to consider quantum superpositions in terms of paraconsistent logics. We claimed that, even though most interpretations of quantum mechanics (QM) attempt to escape contradictions, there are many hints ---coming mainly from present technical and experimental developments in QM--- that indicate it could be worth while to engage in a research of this kind. Recently, Arenhart and Krause \cite{ArenhartKrause14a, ArenhartKrause14b, ArenhartKrause14c} have raised several arguments against this approach. More specifically, taking into account the square of opposition, they have argued that quantum superpositions are better understood in terms of {\it contrariety propositions} rather than {\it contradictory propositions}. In \cite{deRonde14, deRonde15a} we defended the Paraconsistent Approach to Quantum Superpositions (PAQS) and provided arguments in favor of its development. We showed that: {\it i) Arenhart and Krause placed their obstacles from a specific metaphysical stance, which we characterized in terms of what we call the Orthodox Line of Research (OLR).} And also, {\it ii) That this is not necessarily the only possible line, and that a different one, namely, a Constructive Metaphysical Line of Research (CMLR) provides a different perspective in which PAQS can be regarded as a valuable prospect.} Furthermore, we explained how, within the CMLR, the problems and obstacles raised by Arenhart and Krause disappear. 

We characterized the OLR in terms of two main principles: 

\begin{enumerate}
\item {\bf Quantum to Classical Limit:}  The principle that one must find a ``bridge'' or ``limit'' between classical mechanics and QM.
\end{enumerate}

\begin{enumerate}
\item[2.] {\bf Classical Representation of Physics:} The principle that one needs to presuppose the classical representation of physics (which assumes the metaphysical equation: Actuality = Reality) in order to interpret QM.
\end{enumerate}

\noindent More specifically, regarding quantum superpositions, we have showed how the Measurement Problem (MP), one of the main questions imposed by the OLR, has closed the door to a true development of the theory. Given the fact that QM describes mathematically the state of a quantum system in terms of a superposition, the question is why do we observe a single result instead of a superposition of them?\footnote{Given a quantum system represented by a superposition $\sum c_i | \alpha_i \rangle$, when in contact with an apparatus ready to measure, $|R_0 \rangle$, QM predicts that system and apparatus will become ``entangled'' in such a way that the final `system + apparatus' will be described by  $\sum c_i | \alpha_i \rangle  |R_i \rangle$. Thus, as a consequence of the quantum evolution, the pointers have also become ---like the original quantum system--- a superposition of pointers $\sum c_i |R_i \rangle$. This is why the MP can be stated as a problem only in the case the original quantum state is described by a superposition of more than one term.} The MP accepts the fact that there is something very weird about quantum superpositions but, leaving aside their problematic meaning, focuses on the justification of the actualization process. 

Taking distance from the OLR, the CMLR is based on three main presuppositions already put forward and discussed in \cite[pp. 56-57]{deRonde11}.

\begin{enumerate}
\item {\bf Closed Representational Stance:} Each physical theory is closed under its own formal and conceptual structure and provides access to a specific field of phenomena. The theory provides the constraints to consider, explain and understand physical phenomena. {\it It is only the theory which can tell you what can be observed.}

\item {\bf Formalism and Empirical Adequacy:} The formalism of QM is able to provide (outstanding) empirically adequate results. Empirical adequacy determines the success of a theory and not its commitment to a certain presupposed conception of the world. Thus, it seems to us that the problem is not to find a new formalism. On the contrary, the `road signs' point in the direction that {\it we must stay close to the orthodox quantum
formalism}.

\item {\bf Constructive Conceptual Stance:} Since classical physical concepts do not seem well suited to provide a coherent understanding of QM and its phenomena, to learn about what the formalism of QM is telling us about reality we might be in need of {\it creating new physical concepts}.
\end{enumerate}

\noindent Changing the metaphysical standpoint, the CMLR also presents a different questioning which assumes right from the start the need of bringing into stage a different metaphysical scheme to the one assumed by the OLR ---based on the notion of a system  constituted by properties in the actual mode of existence. What is needed, according to this stance, is a radical inversion of orthodoxy and its problems. The CMLR attempts to bring back metaphysical considerations within the analysis of QM taking into account three main ideas: the first is that physical observation is theory-laden and thus always metaphysically founded; the second is that operational counterfactual reasoning about Meaningful Physical Statements (MPS)\footnote{In \cite{deRonde15c} we defined {\bf Meaningful Physical Statements (MPS):} {\it If given a specific situation a theory is capable of predicting in terms of definite physical statements the outcomes of possible measurements, then such physical statements are meaningful relative to the theory and must be constitutive parts of the particular representation of physical reality that the theory provides. Measurement outcomes must be considered only as an exposure of the empirical adequacy (or not) of the theory.}} is the kernel of physical discourse and in consequence cannot be abandoned if we seek to find an objective representation of physical reality; the third and final consideration is that predictions must be necessarily related to the physical representation of reality provided by the theory. \\ 

\noindent {\it {\bf Representational Realist Stance (RRS):} A representational realist account of a physical theory must be capable of providing a physical representation of reality in terms of a network of concepts which relates to the mathematical formalism of the theory and allows to make predictions of definite phenomena.}\\ 

\noindent From this standpoint a coherent interpretation of QM should be able to explain in what sense quantum superpositions, which give rise to different sets of MPS, are part of the physical representation of reality provided by the theory. Instead of considering the MP, which attempts to justify the actual observational realm, we should focus in providing a physical representation of quantum superpositions (see for discussion \cite{deRonde15a})  ---a representation that goes beyond the discourse of measurement outcomes. 

Following this line of research, if we accept the idea that quantum superpositions might exist in a contradictory potential realm ---leaving aside the idea that: Actuality = Reality---, it makes perfect sense to approach QM in terms of a paraconsistent approach and claim that {\it quantum superpositions are contradictory existents}. Following these ideas, and taking as a standpoint a suitable interpretation in terms of the notions of {\it power} and {\it potentia} we present in this paper a paraconsistent formalization of quantum superpositions. The article is organized as follows. In section 1, we discuss the problem of representing quantum superpositions in terms of physical notions. Section 2 analyzes the meaning of modality in QM and puts forward an interpretation in terms of `ontological potentiality'. Section 3, presents the notion of `power' as a real physical existent. In section 4, we analyze the meaning of contradiction in the actual and the potential realms. In section 5  we provide a particular formalization of quantum superpositions in terms of contradictory powers in the potential realm. Finally, section 6 presents the conclusions of the paper.

\section{The Physical Representation of Quantum Superpositions}

The CMLR focuses in the Superposition Problem (SP) which attempts to put forward a physical representation of quantum superpositions leaving aside the famous MP which ---following the OLR--- attempts instead to justify the actual (non-contradictory) realm of existence and concentrates in the analysis of the measurement process.\footnote{{\bf Superposition Problem (SP):} {\it  Given a specific situation in which there is a quantum superposition of more than one term, $\sum c_i \ | \alpha_i \rangle$, and given the fact that each one of the terms relates trough the Born rule to a MPS, the problem is how do we physically represent this mathematical expression, and in particular, the multiple terms?}} The shift in the focus of the problem means at the same time going against a basic tenet taken by many of the founding fathers regarding the presupposed limits of physical representation in quantum theory. Indeed, the idea that quantum superpositions cannot be physically represented was stated already in 1930 by Paul Dirac in the first edition of his famous book, {\it The Principles of Quantum Mechanics}.  

\begin{quotation}
\noindent {\small``The nature of the relationships which the superposition principle requires to exist between the states of any system is of a kind that cannot be explained in terms of familiar physical concepts. One cannot in the classical sense picture a system being partly in each of two states and see the equivalence of this to the system being completely in some other state. There is an entirely new idea involved, to which one must get accustomed and in terms of which {\it one must proceed to build up an exact mathematical theory, without having any detailed classical picture}." \cite[p. 12]{Dirac74} (emphasis added)}\end{quotation}

\noindent Also Niels Bohr was eager to defend the classical physical representation of our world and set the limits of such representation in classical physics itself \cite{BokulichBokulich}. Bohr \cite[p. 7]{WZ} argued that: ``[...] the unambiguous interpretation  of any measurement must be essentially framed in terms of classical physical theories, and we may say that in this sense the language of Newton and Maxwell will remain the language of physicists for all time.'' At the same time he closed any further conceptual development by arguing that ``it would be a misconception to believe that the difficulties of the atomic theory may be evaded by eventually replacing the concepts of classical physics by new conceptual forms.'' Even Schr\"odinger, who was one of the first to see the implications of {\bf the superposition principle}, exposing it through his also famous ``cat experiment'', didn't dare to explore beyond the accepted limits of classical physical reality remaining inside the gates of our classical metaphysical understanding of the world \cite{Schr35}. 

Unfortunately, these ideas have sedimented in the present foundational research regarding QM. Indeed, the strategy of the OLR has been to focus on two main problems which deal with the justification of classical physics. The first is the basis problem which attempts to explain how nature ``chooses'' a single basis ---between the many possible ones--- when an experimental arrangement is determined in the laboratory ---this also relates to the problem of contextuality which we have analyzed in detail in \cite{deRonde11}. The second problem is the already mentioned MP. Given the fact that QM describes mathematically the state in terms of a superposition, the question is why do we observe a single result instead of a superposition of them? The MP focuses on the justification of the actualization process. Taking as a standpoint the single outcome it asks: how do we get to a single measurement result from the quantum superposition?\footnote{The questioning is completely analogous to the one posed by the quantum to classical limit problem: how do we get from contextual weird QM into our classical physical description of the world?} The MP places the outcome in the origin, and what needs to be justified is the already known answer. Contrary to the OLR, our main interest focuses on the SP which stresses the need of providing a physical representation of quantum superpositions (expressed in general by the mathematical expression: $\sum c_i \ | \alpha_i \rangle$). But in order to engage in such a project we first need to go beyond the questioning regarding measurement outcomes. Before we can understand actualization we first need to explain what a quantum superposition {\it is} or {\it represents}. There are multiple ways of interpreting the {\it projection postulate} \cite{RFD14a} and thus there is no self evident path between the superposition and its outcome. Even the OLR will have to recognize that we are in need of providing a proper coherent physical interpretation of quantum superpositions.  

Our research is focused on the hypothesis that quantum superpositions relate to something physically real that exists in Nature, and that in order to understand QM we need to provide a physical representation of such existents. But why do we think we have good reasons to believe that quantum superpositions exist? As clearly expressed by Griffiths \cite[p. 361]{Griffiths02}: ``If a theory makes a certain amount of sense and gives predictions which agree reasonably well with experimental or observational results, scientists are inclined to believe that its logical and mathematical structure reflects the structure of the real world in some way, even if philosophers will remain permanently skeptical." Quantum superpositions are one of the main sources used by present experimental physicists in order to put forward the most outstanding technical and experimental developments of the last decades. Indeed, there are many characteristics and behaviors we have learnt about superpositions: we know about {\it their existence regardless of the effectuation of one of its terms}, as shown, for example, by the interference of different possibilities in {\it welcher-weg} type experiments \cite{ Nature11a, NaturePhy12}, {\it their reference to contradictory properties}, as in Schr\"{o}dinger cat states \cite{Nature07}, we also know about {\it their non-standard route to actuality}, as explicitly shown by the modal KS theorem \cite{DFR06, RFD14a}, and we even know about {\it their non-classical interference with themselves and with other superpositions}, used today within the latest technical developments in quantum information processing \cite{Nature13}. In spite of the fact we still cannot say what a quantum superposition {\it is} or {\it represents}, we must admit that they seem ontologically robust. If the terms within a quantum superposition are considered as quantum possibilities (of being actualized) then we must also admit that such quantum possibilities {\it interact} ---according to the Schr\"odingier equation. It is also well known that one can produce interactions between multiple superpositions (entanglement) and then predict their evolution as well as the {\it ratio} of all possible outcomes. It then becomes difficult not to believe that these terms that `interact', `evolve' and `can be predicted' according to the theory, are not real (in some way). 

Leaving aside these known facts, the orthodox perspective assumed by a numerous group of interpretations seems to disregard the idea that quantum superpositions are real physical existents. Maybe the most known interpretation between them is the so called ``Copenhagen interpretation'' which remains agnostic with respect to the mode of existence of properties {\it prior} to measurement. The same interpretation is endorsed by van Fraassen in his Copenhagen modal variant.\footnote{According to Van Fraassen \cite[p. 280]{VF91}: ``The interpretational question facing us is exactly: in general, which value attributions are true? The response to this question can be very conservative or very liberal. Both court later puzzles. I take it that the Copenhagen interpretation ---really, a roughly correlated set of attitudes expressed by members of the Copenhagen school, and not a precise interpretation--- introduced great conservatism in this respect. Copenhagen scientists appeared to doubt or deny that observables even have values, unless their state forces to say so. I shall accordingly refer to the following very cautious answer as the {\it Copenhagen variant} of the modal interpretation. It is the variant I prefer.''} Much more extreme is the instrumentalist perspective put forward by Fuchs and Peres \cite[p. 1]{FuchsPeres00} who claim that: ``[...] quantum theory does not describe physical reality. What it does is provide an algorithm for computing probabilities for the macroscopic events (``detector clicks") that are the consequences of experimental interventions.'' But even within realistic interpretations there is a complete lack of physical explanation regarding the meaning of quantum superpositions (see for a detailed analysis \cite{deRonde15c}). In Dieks' realistic modal version as well as in Griffiths consistent histories interpretation quantum superpositions are not considered as real physical existents. Given a $\Psi$, only one of its mathematical representations is related to reality (actuality) ---namely, the representation in which $\Psi$ is written as a single term---  while all other representations are considered merely as ``computational tools'' to calculate outcomes. Bohmian versions neglect ``right from the start'' the existence of quantum superpositions and propose instead ---changing the  formalism--- the existence of a quantum field that governs the evolution of particles. One might also argue that some interpretations, although not explicitly, leave space to consider superpositions as existent in a potential, propensity or dispositional realm. The Jauch and Piron School, Popper or Margenau's interpretations, are a clear example of such proposal (see for discussion \cite{deRonde11} and references therein). However,  within such interpretations the collapse is accepted and potentialities, propensities or dispositions are only defined in terms of `their becoming actual' ---mainly because, forced by the OLR, they have been only focused in providing an answer to the MP. In any case, such realms are not articulated beyond their meaning in terms of the actual realm. Only the many worlds interpretation goes as far as claiming that all terms in the superposition are real in actuality. However, the quite expensive metaphysical price to pay is to argue that there is a multiplicity of unobservable worlds (branches) in which each one of the terms is actual. Thus, the superposition expresses the multiplicity of such classical actual worlds.   

Instead of taking one of these two paths which force us either into the abandonment of representation and physical reality or to the exclusive account of physical representation in terms of an Actual State of Affairs we have proposed, through the CMLR, to develop a new path which is focused in developing radically new (non-classical) physical concepts.

\section{Modality in Quantum Mechanics}

QM has been related to modality since its origin, when Max Born interpreted Schr\"odingier's quantum wave function, $\Psi$, as a ``probability wave''. However, it was very clear from the very beginning that the meaning of modality and probability in the context of QM was something completely new. As remarked by Heisenberg himself: 

\begin{quotation}
\noindent {\small ``[The] concept of the probability wave [in QM] was something entirely new in theoretical physics since Newton. Probability in mathematics or in statistical mechanics means a statement about our degree of knowledge of the actual situation. In throwing dice we do not know the fine details of the motion of our hands which determine the fall of the dice and therefore we say that the probability for throwing a special number is just one in six. The probability wave function, however, meant more than that; it meant a tendency for something.'' \cite[p. 42]{Heis58}}\end{quotation}

\noindent Today, it is well known that quantum probability does not allow an interpretation in terms of ignorance ---even though many papers in the literature still use probability uncritically in this way. Instead, as we mentioned above, the quantum formalism seems to imply some kind of weird {\it interaction of possibilities} governed by the Schr\"odingier equation (see for discussion \cite{deRonde15c, deRonde15d, Dieks10}). 

As discussed above, the problems put forward by the OLR do not focus on the development of a new non-classical physical representation of QM and have concentrated their efforts instead, in justifying our classical representation of entities constituted by properties in the actual mode of existence. We are convinced that only when leaving behind the OLR and its problems, one might be able to discuss about a non-classical development of the theory. Of course this implies reconsidering the meaning of existence itself and the abandonment of another presupposed dogma, namely, that existence and reality are represented by actuality either as an observation {\it hic et nunc} (empiricism) or as a mode of existence (realism). Following the CMLR, we believe a reasonable strategy would then be to start with what we know works perfectly well, namely, the orthodox formalism of QM and advance in the metaphysical principles which constitute our understanding of the theory. Escaping the metaphysics of actuality and starting from the formalism, a good candidate to develop a mode of existence that is implicit within the theory itself is of course {\it quantum possibility}. 

In several papers, one of us together with G. Domenech and H. Freytes, have analyzed how to understand possibility in the context of the orthodox formalism of QM \cite{DFR06, DFR08a, DFR08b, DFR09}. From this investigation there are several conclusions which can be drawn. We started our analysis with a question regarding the contextual aspect of possibility. As it is well known, the Kochen-Specker (KS) theorem does not talk about probabilities, but rather about the constraints of the formalism to actual definite valued properties considered from multiple contexts. What we found via the analysis of possible families of valuations is that a theorem which we called ---for obvious reasons--- the Modal KS (MKS) theorem can be derived which proves that quantum possibility, contrary to classical possibility, is also contextually constrained \cite{DFR06}. This means that, regardless of its use in the literature: {\it quantum possibility} is not {\it classical possibility}. In a recent paper, \cite{RFD14a} we have concentrated in the analysis of {\it actualization} within the orthodox frame and interpreted, following the structure, the logical realm of possibility in terms of potentiality.

As we remarked in \cite{deRonde15a}, once we accept we have two distinct realms of existence, namely, an actual realm and a potential realm, we must be careful about the way in which we define {\it contradictions}. Certainly, contradictions cannot be defined in terms of truth valuations in the actual realm, simply because, contradictions cannot be found in actuality ---which is a realm explicitly defined in terms of the Principle of Non-Contradiction (see \cite{deRonde15d}). The physical notion that must be related to quantum superpositions must be, according to our research, an existent in the potential realm ---not in the actual one. The MKS theorem shows explicitly that a quantum wave function implies multiple incompatible valuations which can be interpreted as {\it potential contradictions} \cite{RFD14a}. Our analysis has always kept in mind the idea that classical contradictions, that is, contradictions according to classical logic, are never found in the actual realm. Our attempt is to turn things upside-down: {\it we do not need to explain the actual via the potential but rather, we need to use the actual in order to develop the potential} \cite[p. 148]{deRonde11}. 

Leaving aside the widespread paranoia against contradictions, the PAQS does the job of allowing a further formal development of a realm in which all terms of a superposition co-exist, regardless of actuality. In the sense just discussed the PAQS opens possibilities of development which have not yet been fully investigated. It should be also clear that we are not claiming that all terms in the superposition are actual ---as in the many worlds interpretations--- overpopulating existence with unobservable actual worlds. What we claim is that PAQS opens the door to consider all terms as existent in a realm different of actuality. We claim that just like we need all properties to characterize a physical object, all terms in the superposition are needed for a proper characterization of what exists according to QM. We should stress that each term in a quantum superposition relates directly to a MPS of the theory which can be empirically tested. Contrary to Arenhart and Krause we do not agree that our proposal is subject of Priest' razor: the metaphysical principle according to which we should not populate the world with contradictions beyond necessity \cite{Priest87}. The PAQS does not overpopulate metaphysically the world with contradictions, but rather attempts to take into account all MPS provided by QM. We should not forget that experimentalists are using explicitly such MPS in order to produce in the lab a new technical era.\footnote{Regarding observation it is important to remark that such {\it contradictory potentialities} are observable just in the same way as actual properties can be observed in an object. Potentialities can be observed through actual effectuations in analogous fashion to physical objects ---we never observe all perspectives of an object simultaneously, instead, we observe at most a single set of actual properties.}

The modal interpretation proposed in \cite{deRonde11} attempts to develop ---following the CMLR--- a physical representation of the formalism based on three main notions: the notions of ontological potentiality, power and potentia.\footnote{Different modal approaches might be also considered \cite{daCostaLombardi13} taking into account their specificity with respect to paraconsistency and modality.} The notion of {\it ontological potentiality} has been explicitly developed taking into account what we have learnt from the orthodox formalism about quantum possibility, taking potentiality to its limit and escaping the dogmatic ruling of actuality. Contrary to the teleological notion of potentiality used within many interpretations of QM our notion of ontological potentiality is not defined in terms of actuality. Such perspective has determined not only the need to consider what we call a {\it Potential State of Affairs} ---in analogous fashion to the {\it Actual State of Affairs} considered within classical physical theories---, but also the distinction between {\it actual effectuations}, the effectuation of potentiality in the actual realm, and {\it potential effectuations} which is that which happens in the potential realm regardless of actuality (see for discussion \cite{RFD14a}). Actualization only discusses the actual effectuation of the potential, while potential effectuations remain in the potential realm evolving according to QM. The question we would like to discuss in the following section is: what is that which exists in the potential realm?

\section{Powers as Quantum Physical Existents}

Entities are composed by properties which exist in the actual mode of being. But what is that which exists in the ontological potential realm? We have argued that an interesting candidate to consider is the notion of {\it power}. Elsewhere \cite{deRonde11, deRonde13, deRonde15a, deRonde15d}, one of us has put forward such an ontological interpretation of powers. In the following we summarize such ideas and provide an axiomatic characterization of QM in line with these concepts. 

{\it The mode of being of a power is potentiality}, not thought in terms of classical possibility (which relies on actuality) but rather as a mode of existence ---i.e., in terms of ontological potentiality. {\it Powers are indetermined.} To possess the power of {\it raising my hand}, does not mean that in the future `I {\it will} raise my hand' {\it or} that in the future `I {\it will not} raise my hand'; what it means is that, here and now, I possess a power which exists in the mode of being of potentiality, {\it independently of what will happen in actuality}. Powers do not exist in the mode of being of actuality, they are not actual existents, they are undetermined potential existents. Powers, like classical properties, preexist to observation, but unlike them preexistence is not defined in the actual mode of being as an Actual State of Affairs (ASA), instead we have a {\it potential preexistence} of powers which determines a Potential State of Affairs (PSA). Powers are a conceptual machinery which can allow us to compress experience into a picture of the world, just like entities such as particles, waves and fields, allow us to do so in classical physics. We cannot ``see" powers in the same way we see objects.\footnote{It is important to notice there is no difference in this point with the case of entities: we cannot ``see" entities ---not in the sense of having a complete access to them. We only see perspectives which are unified through the notion of object.} Powers are experienced in actuality through {\it elementary processess}. A power is sustained by a logic of actions which do not necessarily take place, it \emph{is} and \emph{is not}, {\it hic et nunc}. 

A basic question which we have posed to ourselves regards the ontological meaning of a {\it quantum superposition} \cite{deRonde13}. What does it mean to have a mathematical expression such as: $\alpha | \uparrow \rangle + \beta  | \downarrow \rangle$, which allows us to predict precisely, according to the Born rule, experimental outcomes? Our theory of powers has been explicitly developed in order to try to answer this particular question. Given a superposition in a  particular basis, $\Sigma \  c_i | \alpha_i \rangle$, the powers are represented by the elements of the basis, $| \alpha_i \rangle$, while the coordinates in square modulus, $|c_i|^2$, are interpreted as the potentia of each respective power.\footnote{Our approach considers finite bases, the extension to infinite bases is not difficult and will be considered in future works.} {\it Powers can be superposed to different ---even contradictory--- powers.} We understand a quantum superposition as encoding a set of powers each of which possesses a definite {\it potentia}. This we call a {\it Quantum Situation} ($QS$). For example, the quantum situation represented by the superposition $\alpha | \uparrow \rangle + \beta |\downarrow\rangle$, combines the contradictory powers, $| \uparrow \rangle$ and $|\downarrow\rangle$, with their potentia, $|\alpha|^2$ and $|\beta|^2$, respectively. Contrary to the orthodox interpretation of the quantum state, we do not assume the metaphysical identity of the multiple mathematical representations given by different bases \cite{deRondeMassri14}. Each superposition is basis dependent and must be considered as a distinct quantum situation. For example, the superposition $c_{x1} | \uparrow_{x} \rangle + c_{x2} |\downarrow_{x}\rangle$ and the superposition $c_{y1} | \uparrow_{y} \rangle + c_{y2} |\downarrow_{y}\rangle$, which are representations of {\it the same} $\Psi$ and can be derived from one another via a change in basis, are interpreted as two different quantum situations, $QS_{\Psi, B_x}$ and $QS_{\Psi, B_y}$. 

The logical structure of a superposition is such that a power and its opposite can exist at one and the same time, violating the principle of non-contradiction \cite{daCostadeRonde13}. Within the faculty of raising my hand, both powers (i.e., the power `I am {\it able to} raise my hand' and the power `I am {\it able not to} raise my hand') co-exist. A $QS$ is {\it compressed activity}, something which {\it is} and {\it is not} the case, {\it hic et nunc}. It cannot be thought in terms of identity but is expressed as a difference, as a {\it quantum}.

The power interpretation of QM can be condensed in the following eight postulates. 

\begin{enumerate}

{\bf \item[I.] Hilbert Space:} QM is represented in a vector Hilbert space.

{\bf \item[II.] Potential State of Affairs (PSA):} A specific vector $\Psi$ with no given mathematical representation (basis) in Hilbert space represents a PSA; i.e., the definite existence of a multiplicity of powers, each one of them with a specific potentia.

{\bf \item[III.] Actual State of Affairs (ASA):} Given a PSA and the choice of a definite basis $B, B', B'',...,$ etc. ---or equivalently a Complete Set of Commuting Observables (C.S.C.O.)---, a context is defined in which a set of powers, each one of them with a definite potentia, are univocally determined as related to a specific experimental arrangement (which in turn corresponds to a definite ASA). The context builds a bridge between the potential and the actual realms, between quantum powers and classical objects. The experimental arrangement (in the ASA) allows the powers (in the PSA) to express themselves in actuality through elementary processes which produce actual effectuations.

{\bf \item[IV.] Quantum Situations, Powers and Potentia:} Given a PSA, $\Psi$, and the context we call a quantum situation to any superposition of one or more than one power. In general given the basis $B= \{ | \alpha_i \rangle \}$ the quantum situation $QS_{\Psi, B}$ is represented by the following superposition of powers:
\begin{equation}
c_{1} | \alpha_{1} \rangle + c_{2} | \alpha_{2} \rangle + ... + c_{n} | \alpha_{n} \rangle
\end{equation}

\noindent We write the quantum situation of the PSA, $\Psi$, in the context $B$ in terms of the ordered pair, $(| \alpha_{i} \rangle, |c_{i}|^2)$ with $n$ less than or equal to $n$, given by the elements of the basis and the coordinates in square modulus of the PSA in that basis:
\begin{equation}
QS_{\Psi, B} = (| \alpha_{i} \rangle, |c_{i}|^2)
\end{equation}

\noindent The elements of the basis, $| \alpha_{i} \rangle$, are interpreted in terms of {\it powers}. The coordinates of the elements of the bases in square modulus, $|c_{i}|^2$, are interpreted as the {\it potentia} of the power $| \alpha_{i} \rangle$, respectively. Given the PSA and the context, the quantum situation, $QS_{\Psi, B}$, is univocally determined in terms of a set of powers and their respective potentia. (Notice that in contradistinction with the notion of `quantum state' the definition of a `quantum situation' is basis dependent.)

{\bf \item[V.] Elementary Process:} In QM we only observe discrete shifts of energy (quantum postulate). These discrete shifts are interpreted in terms of {\it elementary processes} which produce actual effectuations. An elementary process is the path which undertakes a power from the potential realm to its actual effectuation. This path is governed by the {\it immanent cause} which allows the power to remain preexistent in the potential realm independently of its actual effectuation. Each power $| \alpha_{i} \rangle$ is univocally related to an elementary process represented by the projection operator $P_{\alpha_{i}} = | \alpha_{i} \rangle \langle \alpha_{i} |$.

{\bf \item[VI.] Actual Effectuation of Powers (Measurement):} Powers exist in the mode of being of ontological potentiality. An {\it actual effectuation} is the expression of a specific power in actuality. Different actual effectuations expose the different powers of a given $QS$. In order to learn about a specific PSA (constituted by a set of powers and their potentia) we must measure repeatedly the actual effectuations of each power exposed in the laboratory. (Notice that we consider a laboratory as constituted by the set of all possible experimental arrangements that can be related to the same $\Psi$.)

{\bf \item[VII.] Potentia (Born Rule):} A {\it potentia} is the strength of a power to exist in the potential realm and to express itself in the actual realm. Given a PSA, the potentia is represented via the Born rule. The potentia $p_{i}$ of the power $| \alpha_{i} \rangle$ in the specific PSA, $\Psi$, is given by:
\begin{equation}
Potentia \ (| \alpha_{i} \rangle, \Psi) = \langle \Psi | P_{\alpha_{i}} | \Psi \rangle = Tr[P_{ \Psi} P_{\alpha_{i}}]
\end{equation}

\noindent In order to learn about a $QS$ we must observe not only its powers (which are exposed in actuality through actual effectuations) but we must also measure the potentia of each respective power. In order to measure the potentia of each power we need to expose the $QS$ statistically through a repeated series of observations. The potentia, given by the Born rule, coincides with the probability frequency of repeated measurements when the number of observations goes to infinity.

{\bf \item[VIII.]  Potential Effectuation of Powers (Schr\"odinger Evolution):} Given a PSA, $\Psi$, powers and potentia evolve deterministically, independently of actual effectuations, producing {\it potential effectuations} according to the following unitary transformation:
\begin{equation}
i \hbar \frac{d}{dt} | \Psi (t) \rangle = H | \Psi (t) \rangle
\end{equation}
\noindent While {\it potential effectuations} evolve according to the Schr\"odinger equation, {\it actual effectuations} are particular expressions of each power (that constitutes the PSA, $\Psi$) in the actual realm. The ratio of such expressions in actuality is determined by the potentia of each power.
\end{enumerate}

\noindent According to our interpretation, just like classical physics talks about entities composed by properties that preexist in the actual realm, QM talks about powers with definite potentia that preexist in the (ontological) potential realm, independently of the specific actual context of inquiry or the particular set of actualizations. This interpretational move allows us to define powers independently of the context regaining an objective picture of physical reality independent of measurements and subjective choices. The price we are willing to pay is the abandonment of a metaphysical equation that has been presupposed in the analysis of QM: Actuality = Reality. In the following section, talking into account a typical quantum experience, we discuss in what sense powers are to be considered in terms of contradictory statements in the potential realm.

\section{Potential Contradictions and the Actual Realm}

In order to criticize the notion of contradiction supposedly assumed by the PAQS, Arenhart and Krause \cite{ArenhartKrause14b} presented several conditions which according to their analysis, they claim, become incompatible within our approach. Firstly they propose the following semantic requirement: 

\begin{definition}
{\bf Semantic Requirement (SR):} Contradictory statements [of the language] must have opposite truth values.\end{definition}

\noindent To analyze a specific situation in QM they take the following quantum state: $\alpha  \ | \uparrow_x \rangle +  \  \beta \ | \downarrow_x \rangle$ and claim that: ``The statements corresponding to properties represented by $| \uparrow_x \rangle \langle \uparrow_x |$ and $| \downarrow_x \rangle \langle \downarrow_x |$ must have opposite truth values.'' [{\it Op. cit.}, p. 2] They then proceed to consider two property ascriptions, the minimal property ascription condition. 

\begin{definition}
{\bf Minimal Property Ascription (MPA):} If a system is in an eigenstate of an operator with eigenvalue $v$, then the system has the qualitative property corresponding to such value of the observable.
\end{definition}

\noindent And the paraconsistent property ascription: 

\begin{definition}
{\bf Paraconsistent Property Ascription (PPA):} When in a superposition [of the type $\alpha  \ | \uparrow_x \rangle +  \  \beta \ | \downarrow_x \rangle$], the system does have the properties related to the vectors forming the superposition, and they are contradictory.
\end{definition}

Arenhart and Krause then argue that: ``when the conditions for application of the minimal principle [MPA] are met, both states have opposite truth values. But the job is still not done: we must still grant that one of those propositions must always be the case (being so that the other one will be false), as the semantic requirement [SR] for a contradiction seems to demand.'' [{\it Op. cit.}, p. 5] Their conclusion is then the following: 

\begin{quotation}
\noindent {\small``[...] it seems that the semantic requirement [SR] that one of the two terms in a superposition must always be the case (so that we can have a contradiction) is in fact in conflict with the paraconsistent property attribution [PPA] principle. For the latter principle to apply, in the case of a superposition, both `up' and `down' would {\it have to be the case simultaneously.} Recall what happens in the case of the two slit or Schr\"odinger's cat: according to this proposal [PAQS], the particle must go by both slits, the cat must be dead and alive. So, there cannot be alternate truth values in this case, for both must be simply true. So, there is a conflict of the paraconsistent property attribution [PPA] principle with the very requirement that the vectors in a superposition stand for contradictory properties, at least according to the usual semantic requirements [SR] related to contradictions, as it appears in the traditional analysis of this concept. It seems that one cannot have both the claim that $u_x$ [$| \uparrow_x \rangle \langle \uparrow_x |$] and $d_x$ [$| \downarrow_x \rangle \langle \downarrow_x |$] are contradictory and the claim that a superposition involves contradictions, as supplied by the paraconsistent property attribution [PPA] principle. As it stands, it seems, these demands are incompatible.'' [{\it Op. cit.}, p. 5]}
\end{quotation}

Let us analyze the presuppositions for the argument to stand. Both conditions, SR and MPA, imply an analysis either in terms of actuality or actualization. But as we mentioned above, the notion of contradiction that we mean to put forward must consider a realm of potentiality truly independent of actuality. If the superposition (as a physical notion) exists in potentiality, such conditions cannot be taken into account for they implicitly assume that what is found out in actuality must be directly referred to superpositions irrespectively of their mode of existence. However, if the mode of existence to which such conditions make reference are considered, there is plenty of room to take them into account within the PAQS. 

An analogy with the measurement of a power can show us why the argument of Krause and Arenhart does not follow irrespectively of the metaphysical considerations of the subject under study. Let us, for the sake of the argument, admit that powers ---which, as discussed above, are not entities--- exist in a potential realm. There exist contradictory powers such as `putting my hand up' or `putting my hand down' (following the PPA). If I `put my hand up' (in actuality) then everyone who is looking will learn that I posses such a specific power (as demanded by the MPA), and at the same time, everyone will have observed that (in actuality) I did not `put my hand down' (only one of the two possibilities will be `true' in the actual realm as required by the SR) ---vice versa, when I `put my hand down'. The expression of a power in actuality exposes its existence in exactly the same way we can only observe an object when light shines upon it. 

In order to make even more clear our proposal let us consider some questions. First question: does the measurement of a power involve a collapse? The answer is no: the expression of a power in terms of its actual effectuation does not imply that the power has been destroyed nor that it has ceased to exist. The fact that I `put my hand up' in actuality does not imply in any way that I will cease to have this power in the future or that the power has suffered some kind of deterioration. The power remains in the potential realm capable of producing new actual (and potential) effectuations. Second question: do we need actuality to claim that a power exists? The answer is no: I could choose not to raise my hand but nonetheless still claim that the power exists ---in the same way that when I close my eyes I can still claim that the table in front of me exists. According to our RRS, physical reality is completely independent of particular human observations. This simple example attempts to show that if one moves away from the metaphysics of `entities' and `properties', there are different ways to think about physical reality. Entities and properties in the actual mode of being might not be the end of the road.

What is at stake in QM is the meaning itself of existence ---QM does not seem to make reference to an ASA---, this is why we have insisted that we are not committed necessarily to the equation implied by the OLR, `Actuality = Reality'. SR and MPA are not {\it necessary conditions} for every interpretation of QM that we can think of. They already imply a metaphysical stance in which reality is conceived only in terms of entities constituted by properties in the actual mode of existence. But, as we have shown above, if we include the mode of existence of potentiality within the conditions themselves, we can certainly provide a consistent account of SR, MPA and PPA within the PAQS. By claiming that PPA refers to the potential realm while MPA refers to actualization ---something we have discussed in detain in \cite{deRonde14, deRonde15a}--- and SR to the actual realm, all conditions are met by the PAQS and the problems raised by Arenhart and Krause disappear. 

The misinterpretation put forward by  Arenhart and Krause with respect to the PAQS is clarified once we recognize that the classical notion of truth is directly related to the actual realm and that the introduction of a potential realm ---truly independent of actuality--- imposes the need of also developing a (non-classical) potential notion of truth. It is exactly this which will be discussed in the following section.

\section{Truth in the Potential Realm}

An important point which needs to be clearly understood is the fact that the introduction of an ontological realm independent of actuality implies the need of also introducing a different notion of truth to the classical one ---related to the correspondence of a sentence to an ASA. Indeed, the statements that we must consider in relation to quantum superpositions do not fulfill the constraints imposed by the classical notion of truth related to the metaphysical realm of actuality. 

We define potential truth, which we call {\it p-truth}, in relation to the set of powers that {\it potentially preexist} in a PSA. This means we have a set of powers $\{ {\cal P}_{i} \}$ that, given a PSA, are {\it p-true} (see for a definition \cite{RFD14a}). Given a specific choice of a basis, such powers ${\cal P}_{i}$ become actualizable. The choice of the basis or context is now a gnoseological choice, one that has the purpose of analyzing in different ways the particular PSA ---this is completely analogous to the analysis of an ASA though different experimental set ups. Notice that within our scheme, we recover an objective account of quantum contextuality. In the same way as in classical physics we must observe a physical object in different ways in order to lear about its actual properties, quantum contexts become particular exposures of a {\it potentially preexistent} state of affairs. While in classical physics an ASA is composed by objects with {actually preexistent} properties, in QM a PSA is composed by {\it potential preexistent} powers, each one them possessing a definite potentia. While classical statements of the type `the system S is composed by the property p' are either {\it true} or {\it false}, the quantum statements `the quantum situation $QS$ is composed by the power ${\cal P}_{k}$  with potentia ${\bf p_k}$' has two levels of analysis: it might be {\it p-true} or {\it p-false} in relation to a specific PSA, and it might be {\it true} or {\it false} in relation to a particular actualization. While the measurement of a property requires a {\it yes-no experiment}, the measurement of a power and its respective potentia requires a {\it statistical experiment} (see for discussion \cite{deRonde15c}). As analyzed in \cite{deRonde15d}, the Born rule becomes ---within this interpretation--- an objective (probabilistic) account of the (ontological) potentia of a power. 

We might remark that the main misunderstanding produced in the analysis by Krause and Arenhart comes from applying the classical notion of truth (which is directly related to actuality) in order to discuss the notion of contradictory power. Given a PSA and a $QS_{\Psi, B}$ in which we have the the following quantum superposition: $\alpha  \ | \uparrow_x \rangle +  \  \beta \ | \downarrow_x \rangle$. Both powers $ | \uparrow_x \rangle$ and  $| \downarrow_x \rangle$ are potentially preexistent, i.e. the statements `the power $| \uparrow_x \rangle$ exists in the potential realm' and `the power $| \downarrow_x \rangle$ exists in the potential realm'  are both {\it p-true} but not {\it true}. The powers cannot be accounted for by a single {\it yes-no experiment}. As a matter of fact, if we measure the superposition we know that only one power will become actual. One of the powers will be {\it true} while the other will be {\it false}. 

Our definition of {\it potential contradiction} is related to the actualization of the potential  realm. In our example the {\it p-true} powers are contradictory only with respect to the actual realm. Recalling the definition of the square of opposition: \\

\noindent {\bf Contradictory Propositions:} $\alpha$ and $\beta$ are {\it contradictory} when both [potential powers] cannot be [actually] true and both cannot be [actually] false.\\

\noindent From the above analysis, it is the uncritical use of the classical notion of truth ---which relates to an actualist metaphysical perspesctive--- which is responsible for many misunderstandings regarding quantum superpositions and their meaning in terms of physical reality. Krause and Arenhart argue with respect to Schr\"odinger's cat that, ``according to [the PAQS], [...] the cat must be dead and alive. So, there cannot be alternate truth values in this case, for both must be simply true.'' The resolution to the paradox comes from the distinction of both realms and from the proper definition of the specific notion of truth used in each one of the realms considered. Both powers in a $QS$ are {\it p-true}, but only one of them will be {\it true}. This shows clearly that there is no conflict with the PPA in the potential realm nor with the SR in the actual realm. 

In the following section, continuing with the PAQS proposal \cite{daCostadeRonde13}, we provide a paraconsistent formalization of the notion of power that is physically coherent with our previous interpretation of quantum superpositions.

\section{Formalizing Contradictory Powers in the Potential Realm}

In \cite{daCostadeRonde13} we delineated a paraconsistent logical foundation $ZF_1$ for non-relativistic QM. Let us suppose that a theory $T$ is based on a logic $L$ and that in $L$ there is a symbol $\neg$ that corresponds to negation. $T$ is called inconsistent if there exists at least a formula $A$ such that $A$ and $\neg A$ are both theorems of $T$; otherwise, $T$  is called consistent. T is said to be trivial if all its formulas are theorems; otherwise, T is said to be non trivial. $L$ is paraconsistent if it can be the underlying logic of theories that are both inconsistent and non trivial. Clearly, if $L$ has, say, two negations, a theory, based on $L$, may be consistent with reference  to one of the negations, but not in connection with the other. (Here a monadic propositional symbol $\neg$ is called a weak negation if it has a significant number of formal properties of classical negation, and a pair of formulas  $A$  and $\neg A$,
being theorems of a theory $T$, whose underlying logic is $L$, do not necessarily cause the triviality of  $T$.) If $A$  and  $\neg A$ are two formulas, where $\neg$ is a negation, they are called $\neg$-contradictory. When there is only one negation $\neg$, we simply say that
$A$  and  $\neg A$ are contradictory formulas. When $L$ (and in consequence $T$)  possesses a conjunction,  $\wedge$, with the usual properties, $A$  and $\neg A$ are
theorems of $T$ if, and only if,  $A \wedge \neg A$ is also a theorem of $T$. Contradictions are formulas of form  $A \wedge \neg A$. If $L$ has two or more negations, we have
different kinds of contradictions.

The system $ZF_1$ of \cite{daCostadeRonde13} contains two types of negation. The weak (primitive) negation $\neg$ and the strong (defined) negation $\neg^{*}$. Therefore, any theory $T$, having $ZF_1$ as its underlying logic, may be weakly inconsistent, but not strongly inconsistent, etc. In particular, we have in $ZF_1$:

$$(A \wedge \neg A) \rightarrow B$$

\noindent where $A$ and $B$ are formulas, is not a theorem, but 

$$(A \wedge \neg^{*}A) \rightarrow B$$

\noindent is probable. 

Taking into account the axioms and rules of $ZF_1$, any  strong contradiction, provable in $ZF_1$, causes its triviality. However, weak contradictions do not, in general, imply triviality. Another important point is that classical elementary logic and usual set theory $(ZF)$ are both obtainable inside $ZF_1$  if strong negation, $\neg ^*$, is employed. So, it is easy to prove that any contradiction that doesn't entail the triviality of $ZF_1$ has to be a weak contradiction. These results are consequences of the fact that $ZF_1$ was conceived to contain, in a certain sense classical logic, set theory and traditional mathematics, and, simultaneously, to allow for certain types of contradictions. In effect, as already referred to in \cite{daCostadeRonde13}, $ZF_1$ constitutes a paraconsistent set theory in which theories may be constructed that are weakly inconsistent though non-trivial. ($ZF_1$ seems to be strongly consistent, and it  is possible to show that if this statement is true, then the usual $ZF$ would also be consistent.)
     
Apparently, $ZF_1$ is not only weakly consistent but also strongly consistent. Weakly inconsistencies can only be derived in a theory  $T$, built on $ZF_1$, if they are originated by the specific postulates of $T$. So, our main task is to show how some of such contradictions do occur in $ZF_1$ with the postulates of $T$ in case that $T$ is QM. Therefore, we introduce in this discipline new axioms which reflect  the existence of some weak contradictions inherent in QM. However, we must be careful, because $ZF_1$ contains as subsystems, elementary classical logic and set theory. Taking into account the preceding considerations, we are going to show how weak contradictions naturally appear in $QM_1$, that is, in non relativistic QM founded on $ZF_1$.\\

We shall take into account, in what follows, a fixed PSA, $\Psi$, and a definite $QS_{\Psi, B}$,  given by the representation of $\Psi$ in a specific basis B. In $QM_1$ we postulate the existence of powers. So we adopt the following:

\begin{postulate}
There exist a non-empty set of objects ${\cal P}$ called powers. 
\end{postulate}

\begin{definition}
A quantum event is an ordered pair ({\bf A}, $\Delta$), where {\bf A} is an hermitian operator and $\Delta$ is a Borel set of reals.
\end{definition}

\begin{definition}
A quantum statement is an assertion according to which the power  ${\cal P}_{k}$ has the potentia ${\bf p_{k}}$ when the PSA is given by the wave function $\Psi$. When the power ${\cal P}_{k}$ is part of the PSA we say that it is a p-true quantum statement. Given a `quantum event $k$' we say that the correspondent power ${\cal P}_{k}$ is true.
\end{definition}

\begin{postulate}
To any p-true quantum statement there corresponds a power with a definite potentia and conversely. The measure of the potentia is given via the Born rule and coincides with the quantum probability $\langle \Psi | P_{\alpha_{i}} | \Psi \rangle = Tr[P_{ \Psi} P_{\alpha_{i}}]$.
\end{postulate}

\begin{definition}
Given the true quantum statement s, the corresponding power is denoted by ${\cal P}_s$. Each power has a corresponding potentia ${\bf p}_s$.
\end{definition}

\noindent The preceding postulates allows us to extend to the set of powers ${\cal P}$ with potentia {\bf p}, the propositional syntactical structure of the sentences of ${\cal C}_1$, which is the propositional logic underlying $ZF_1$. In the reminder of this article, we shall treat only the case in which the basis of the Hilbert space has cardinal 2. But the extension to all cases is not difficult. QM implies that in certain situations such as a typical Stern Gerlach experiment one needs to consider contradictory powers. To obtain contradictions or inconsistencies in $QM_1$, one has to postulate, directly, their existence. 

\begin{postulate}
If we have a $QS_{\Psi, B}$ represented by $c_1 | \uparrow \rangle + c_2 | \downarrow \rangle$ (which can be also noted $c_1 {\cal P}_{\uparrow} + c_2 {\cal P}_{\downarrow}$) then the quantum statements `the power ${\cal P}_{\uparrow}$ is in the $QS_{\Psi, B}$ and its potentia has the value ${\bf p_{\uparrow}} = |c_1|^2$'  and `the power ${\cal P}_{\downarrow}$ is in the $QS_{\Psi, B}$ and its potentia has the value ${\bf p_{\downarrow}} = |c_2|^2$' are both {\it p-true}. If $| \uparrow \rangle$ is contradictory to $| \downarrow \rangle$ (e.g., as in a SG experiment) then ${\cal P}_{\uparrow} \wedge \neg {\cal P}_{\uparrow}$ and ${\cal P}_{\downarrow}  \wedge \neg {\cal P}_{\downarrow}$, where $\neg$ is the weak negation of $ZF_1$.
 \end{postulate}

\noindent In consequence, the superposition entails for powers ${\cal P}_{\bf A}$ with potentia, ${\bf p_ A}$ ($0 <{\bf p_ A} < 1$), the existence of weakly contradictory powers in the just mentioned case, with the potencies $|c_1|^2$ and $|c_2|^2$. We could reinforce this last postulate 6.6 assuming, in addition, that ${\cal P}_\uparrow \rightarrow \neg {\cal P}_\uparrow$ and ${\cal P}_\downarrow \rightarrow \neg {\cal P}_\downarrow$.

According to the spectral theorem, to every quantum event $({\bf A}, \Delta)$ corresponds a projection operator $P^{A}_{\Delta}$, satisfying the following condition: the probability that a measurement of A in the PSA $\Psi$ yields a value in {\bf A} is $\langle \Psi | P^{{\bf A}}_{\Delta} | \Psi \rangle$. Therefore, we have: 

\begin{prop}
Since our underlying logic (and set theory) is $ZF_1$, that contains $ZF$ when the strong negation is employed in the role of classical negation, then the set of powers  (and that of events) has the structure of the set of all  subspaces of a Hilbert space.
\end{prop}

\noindent Therefore, 

\begin{prop}
In $ZF_1$, the sets of quantum propositions and the set of associate powers constitute an orthomodular lattice.
\end{prop}

In a future paper we attempt to formally account for the incompatibility of multiple quantum situations and powers.

\section{Conclusions}

In this paper we have provided a paraconsistent formalization of the power interpretation of QM \cite{deRonde13, deRonde15a, deRonde15d}. We should remark that this proposed interpretation of the PAQS is one in between many possible interpretations that could use the PAQS for their own theoretical and conceptual developments. Both formal and conceptual developments open new doors to understand physical theories. We should not stick dogmatically to presupposed notions and formal schemes simply because they have worked fairly well in the past (e.g., the notion of particle, actual property, etc.). As remarked by Einstein, this can lead to the impossibility of true developments within science itself: 

\begin{quotation}
\noindent {\small``Concepts that have proven useful in ordering things easily
achieve such an authority over us that we forget their earthly
origins and accept them as unalterable givens. Thus they come to be
stamped as `necessities of thought,' `a priori givens,' etc. The
path of scientific advance is often made impossible for a long time
through such errors.'' \cite[p. 102]{Einstein16}}
\end{quotation}

\noindent We should remain open to new developments that can allow us to discuss QM from new original perspectives. The PAQS attempts to advance exactly in this direction. 

\section*{Acknowledgments} The authors would like to thank G. Domenech, H. Freytes, D. Krause and J. Arenhart for many discussions regarding the meaning of quantum superpositions. This work was partially supported by the following grants: FWO project G.0405.08, FWO-research community W0.030.06 and CONICET RES. 4541-12.

\end{document}